\documentstyle[preprint,floats,prl,aps,epsf]{revtex}

\begin{document}

\preprint{\hbox {June 2003} }

\draft
\title{Robustness and Predictivity of 4 TeV Unification}
\author{\bf Paul H. Frampton, Ryan M. Rohm and Tomo Takahashi}
\address{Department of Physics and Astronomy,\\
University of North Carolina, Chapel Hill, NC  27599.}
\maketitle
\date{\today}

\begin{abstract}
The stability of the predictions of two of the standard model
parameters, $\alpha_3(M_Z)$ and $\sin^2 \theta(M_Z)$,
in a $M_U \sim 4$ TeV unification model is examined.
It is concluded that varying the unification scale
between $M_U \simeq 2.5$ TeV and $M_U \simeq 5$ TeV
leaves robust all predictions within reasonable bounds.
Choosing $M_U = 3.8 \pm 0.4$ TeV gives, at lowest order,
accurate predictions at $M_Z$.
The impact of threshold effects on unification
depends on the spectrum of states beyond the standard model.
\end{abstract}

\newpage

\bigskip
\bigskip

\noindent {\it Introduction}

\bigskip

\noindent One of the principal motivations for extending
the standard model is the GUT gauge hierarchy between the weak
scale and the grand unification or GUT scale.
A related concern, not addressed here, is the Planck hierarchy 
between the weak scale and the Planck scale; the model
we consider has flat spacetime, vanishing Newton's constant
and infinite Planck scale.

The most popular solution of the GUT hierarchy is low-energy
supersymmetry\cite{DG,DRW,ADF,ADFFL} where the three gauge couplings
$\alpha_i(\mu)$ (i = 1,2,3) run logarithmically 
from $\mu = M_Z \sim 91$ GeV, where they are known, up to 
$M_{GUT} \sim 2 \times 10^{16}$ GeV, where they coincide 
with impressive accuracy.

In a recently-proposed model\cite{PHF02}, grand unification occurs
differently. The three couplings run from $\mu = M_Z$
up to a lower unification scale $M_U \sim 4$ TeV,
at which scale the theory is embedded in a larger
gauge group $G \equiv SU(3)^{12}$. The $SU(3)$ gauge couplings
$\alpha_j(\mu)$ (j=1-12) are all equal at $\mu = M_U$.
The embedding of the standard model gauge group
in the larger gauge group $G$ provides a group-theoretical
explanation for the different values of $\alpha_i(M_U)$.

This low-scale unification model also has a top-down inspiration
from string theory through the AdS/CFT correspondence
\cite{M,GKP,W} arising from consideration of a Type IIB 
superstring in d = 10 dimensional
spacetime compactified on $AdS_5 \times S^5$.
Using a finite group 
$\Gamma = Z_{12}$ in an abelian orbifold $AdS_5 \times S^5/\Gamma$ 
gives a quiver gauge theory\cite{PHF99} with gauge group
$SU(N)^{12}$ either with no supersymmetry
${\cal N} = 0$ \cite{PHF02} or
with ${\cal N} = 1$ supersymmetry\cite{FK}

Several issues were left open in \cite{PHF02}:
robustness of the predictions under variations
of the scale $M_U$ (conversely, the accuracy of the
predictions at $\mu = M_Z$); 
the size of flavor-changing effects, and
the consistency of the additional states around $M \sim M_U$
with constraints imposed by precision low-energy data.
In this article we shall address all of these issues.

\newpage

\bigskip
\bigskip

\noindent {\it Robustness of Predictions to Variation in $M_U$}

\bigskip

\noindent The calculations of \cite{PHF02} were done in the one-loop
approximation to the renormalization group equations
without threshold effects.
Because the couplings remain weak this can be self-consistent
provided the masses of the new states 
in the model are sufficiently close to $M_U$.
Other corrections due to non-perturbative effects, and the effects of
large extra dimensions, are outside of the scope of this paper. 
In one sense the robustness of this TeV-scale
unification is almost self-evident, in that it follows from the weakness
of the coupling constants in the evolution from $M_Z$ to $M_U$.
That is, in order to define the theory at $M_U$,
one must combine the effects of
threshold corrections ( due to O($\alpha(M_U)$)
mass splittings )
and potential corrections from redefinitions
of the coupling constants and the unification scale.
We can then {\it impose} the coupling constant relations at $M_U$
as renormalization conditions and this is valid
to the extent that higher order corrections do
not destabilize the vacuum state.

We shall approach the comparison with data in two
different but almost equivalent ways. The first
is ``bottom-up", where we use as input the
requirement that the values of $\alpha_3(\mu)/\alpha_2(\mu)$ and
$\sin^2 \theta (\mu)$ are expected to be $5/2$
and $1/4$, respectively, at $\mu = M_U$.
Using the experimental ranges allowed for
$\sin^2 \theta (M_Z) = 0.23113 \pm 0.00015$,
$\alpha_3 (M_Z) = 0.1172 \pm 0.0020$ and
$\alpha_{em}^{-1} (M_Z) = 127.934 \pm 0.027$
from \cite{PDG} we have plotted in Figure 1
the values of $\sin^2 \theta (M_U)$
(vertical axis) and $\alpha_3 (M_U) / \alpha_2(M_U)$
(horizontal axis) for a range of $M_U$ between 1.5 TeV
and 8 TeV.
Allowing a maximum discrepancy of $\pm 1\%$ in
$\sin^2 \theta (M_U)$ and 
$\pm 4\%$ in $\alpha_3 (M_U) / \alpha_2 (M_U)$
as reasonable estimates of corrections, we deduce that
the unification scale $M_U$ may vary
between 2.5 TeV and 5 TeV. Thus the theory is
robust in the sense that uncertainty in the
renormalization group equations does not effect the
existence of unification.

\newpage

\bigskip
\bigskip

\noindent {\it Accuracy of Predictions at $\mu = M_Z$}

\bigskip

\noindent Alternatively, to test of predictivity 
we fix the unification values at $M_U$ of
$\sin^2 \theta(M_U) = 1/4$ and $\alpha_3 (M_U) / 
\alpha_2 (M_U) = 5/2$ and compute the
resultant predictions at the scale $\mu = M_Z$.
The results are shown for $\sin^2 \theta (M_Z)$
in Fig. 2 with the allowed range\cite{PDG}
$\alpha_3 (M_Z) = 0.1172 \pm 0.0020$. The precise
data on $\sin^2 (M_Z)$ are indicated in Fig. 2 
demonstrating that the model makes correct
predictions for $\sin^2 \theta (M_Z)$.
Similarly, in Fig 3, there is a plot of the
prediction for $\alpha_3 (M_Z)$ versus
$M_U$ with $\sin^2 \theta(M_Z)$ held
within the allowed empirical range. 
The two quantities plotted in Figs 2 and 3
are consistent for similar ranges of $M_U$:
both $\sin^2 \theta(M_Z)$ and $\alpha_3(M_Z)$ 
are within the empirical limits
if $M_U = 3.8 \pm 0.4$ TeV.

\newpage

\bigskip
\bigskip

\noindent {\it Precision Electroweak Data}

\bigskip

\noindent The model has many additional gauge bosons
at the unification scale, including neutral $Z^{'}$'s
and charged $W's$, which could mediate flavor-changing processes
on which there are strong empirical upper limits.
The lower bound on a $Z^{'}$
coupling like the standard $Z$
is $M(Z^{'}) < 1.5$ TeV \cite{PDG} which is below the $M_U$ 
values considered here; however, the couplings of the other $SU(3)$
gauge groups associated with $SU(3)_W$ have a coupling generically stronger
by a factor $4$ requiring that $M(Z^{''}) < 6 TeV$ and hence a real
danger of too-strong FCNC. This is, in our view, the tightest
constraint on the viability of such conformality models.
Full analysis requires commitment to
a specific identification
of quark flavors in the quiver diagram .

Since there are many new states predicted 
at the unification scale $\sim 4$ TeV, there is, in addition, a potential 
of being ruled out by other precision low energy data, as
conveniently studied in terms
of the parameters $S$ and $T$ introduced in \cite{Peskin},
designed to measure departure from the predictions
of the standard model.
Concerning $T$, if the new $SU(2)$ doublets are
mass-degenerate and hence do not violate a custodial
$SU(2)$ symmetry, they do not contribute $T$.
This provides a constraint on the spectrum
of new states.
According to \cite{Peskin}, a multiplet of degenerate
heavy chiral fermions gives a contribution to $S$:

\begin{equation}
S = C \sum_i \left( t_{3L}(i) - t_{3R}(i) \right)^2 / 3 \pi
\label{capitalS}
\end{equation} 
where $t_{3L,R}$ is the third component of weak isospin
of the left- and right- handed component of
fermion $i$ and $C$ is the number of colors.
In the present model, the additional fermions are non-chiral
and fall into vector-like multiplets and so do not
contribute
to $S$.
Provided that the extra isospin multiplets 
at the unification scale $M_U$ are sufficiently
mass-degenerate, therefore, there is no conflict
of chiral fermions with precision data at low energy.

For contribution of new gauge bosons, we refer to the analysis in \cite{FH}.
In the limit where the bilepton gauge bosons are degenerate
$M_{++} = M_{+}$
the contribution to $S$ vanishes except for the subtlety of the pinch
contribution. From the formula presented in \cite{FH}
we find ($\left. S \right\vert_P$ is the pinch contribution):
\begin{equation}
S = S_0 + \left. S \right\vert_P
\label{S}
\end{equation}
The first term in Eq.(\ref{S}) is explicitly:
\begin{eqnarray}
S_0 &=& - 16 \pi {\rm Re} \frac{ \Pi^{3Y}(m_Z^2) - \Pi^{3Y}(0) }{m_Z^2}
\nonumber\\
&=& \frac{9}{4\pi}
\Biggl[
  \ln \frac{M_{++}^2}{M_+^2}
  + \frac{2}{m_Z^2}
  \left(
    M_{++}^2 \bar{F}_0(m_Z^2,M_{++},M_{++})
    - M_{+}^2 \bar{F}_0(m_Z^2,M_{+},M_{+})
  \right)
\nonumber\\
&& \qquad
  + \frac{4}{3}
  \left(
    \bar{F}_0(m_Z^2,M_{++},M_{++})
    - \bar{F}_0(m_Z^2,M_{+},M_{+})
  \right)
\nonumber\\
&& \qquad
  - 2
  \left(
    \bar{F}_3(m_Z^2,M_{++},M_{++})
    - \bar{F}_3(m_Z^2,M_{+},M_{+})
  \right)
\Biggr]
\ ,
\end{eqnarray}
in which $\bar{F}_{0,3}$ are given by:
\begin{eqnarray}
\lefteqn{\bar{F}_0(s,M,m) =
  \int^1_0 dx\, \ln \left( (1-x)M^2 + x m^2 - x(1-x)s \right)
  - \ln M m }
\nonumber\\
&=&
\begin{array}{l}
  \frac{2}{s} \sqrt{(M+m)^2-s} \sqrt{s-(M-m)^2}
  \tan \sqrt{\frac{s-(M-m)^2}{(M+m)^2-s}}
  + \frac{M^2-m^2}{s} \ln \frac{M}{m} - 2 , 
\end{array}
\end{eqnarray}
and
\begin{eqnarray}
\lefteqn{
  \bar{F}_3(s,M,m) =
  \int^1_0 dx\, x(1-x)\ln \left( (1-x)M^2 + x m^2 - x(1-x)s \right)
  - \frac{1}{6}\ln M m
}
\nonumber\\
&=&
\frac{1}{6} \left[
  1 + \frac{M^2+m^2}{s} - \frac{2(M^2-m^2)^2}{s^2}
\right]
\bar{F}_0(s,M,m)
\nonumber\\
&& \quad
-\frac{1}{6} \left( 1 - \frac{2(M^2+m^2)}{s} \right)
\frac{M^2-m^2}{s} \ln \frac{M}{m} + \frac{1}{18}
- \frac{(M^2-m^2)^2}{3s^2} . 
\end{eqnarray}
The second term in Eq.(\ref{S}) is:
\begin{eqnarray}
\left. S \right\vert_P &=& \frac{1}{\pi}
\Biggl[
  3 \ln \frac{M_{++}^2}{M_{+}^2}
  + 2 ( 1 + 2 \sin^2\theta_W ) \bar{F}_0(m_Z^2,M_{++},M_{++})
\nonumber\\
&& \qquad
  - ( 1 - 4 \sin^2\theta_W ) \bar{F}_0(m_Z^2,M_{+},M_{+})
\Biggr]. 
\end{eqnarray}
From these equations, we find that the contributions
of gauge bosons to $S$ are suppressed by
$(M_Z/M_U)^2 \sim 10^{-4}$ and so even for many
such new gauge bosons the contribution
to $S$ is acceptably small provided the
$SU(2)$ doublets are adequately degenerate.

\newpage

\bigskip
\bigskip

\noindent {\it Threshold Effects}

\bigskip

In the above analysis we have assumed all the new states beyond the standard
model are essentially mass degenerate at $M_U$. More realistically, a subset
of the new states may lie below $M_U$ and consequently effect the running
of the couplings $\alpha_{3_c,2_L,Y}$ because of changes
in the corresponding renormalization group $\beta-$functions.

For the chiral fermions there are 48 bifundamental representations
under $SU(3)^{12}$, some of which are in the light sector of the
standard model, but most of which are heavy. We may label them by
their transformation properties under $SU(3)_C \times SU(3)_W \times
SU(3)_H$ and they are shown in Table 1. In the
normalization\cite{ADFFL,GFT} of the $\beta-$function a factor
$g^3/(16\pi^2)$ has been absorbed so that in the three-family minimal
standard model (MSM) with one Higgs doublet: $\beta^{MSM}_{3C} = -7,
\beta^{MSM}_{2L}= -19/6$ and $\beta^{MSM}_{Y}= 41/6$.  All the
$\Delta\beta$ entries in Table 1 are necessarily positive.  Note that
$Y = (2/\sqrt{3})( T_{8W} - T_{8H})$ with 
$T_8 = {\rm diag}(1/\sqrt{12})(1,1,-2)$.

\[
\begin{tabular}{|l|c|c|c|}  \hline \hline
FERMION MULTIPLET  & $\Delta \beta_C$ 
                   & $\Delta \beta_{2L}$ & $\Delta \beta_Y$   \\
\hline \hline
CC: $({\bf 3},\bar{\bf 3})_C$ 
                   & 2 & 0 & 0      \\  \hline
CW: $5({\bf 3}_C,\bar{\bf 3}_W) + 2(\bar{\bf 3}_C,{\bf 3}_W)$ 
                   & 7 & 7  & 28/3  \\
~~~~~~ $2({\bf 3}_C,\bar{\bf 3}_W) + 2(\bar{\bf 3}_C,{\bf 3}_W)$ 
                   & 4 & 4  & 16/3  \\  \hline
CH: $2({\bf 3}_C,\bar{\bf 3}_H) + 5(\bar{\bf 3}_C,{\bf 3}_H)$ 
                   & 7 & 0  & 28/3  \\
~~~~~~ $2({\bf 3}_C,\bar{\bf 3}_H) + 2(\bar{\bf 3}_C,{\bf 3}_H)$ 
                   & 4 & 0  & 16/3  \\  \hline
WW: $9({\bf 3},\bar{\bf 3})_W$  
                   & 0 & 9  & 12    \\  \hline
HH: $9({\bf 3}, \bar{\bf 3})_H$  
                   & 0 & 0  & 12    \\  \hline
WH: $9({\bf 3}_W,\bar{\bf 3}_H) + 6(\bar{\bf 3}_W,{\bf 3}_H)$ 
                   & 0 & 15 & 40    \\
~~~~~~ $6({\bf 3}_W,\bar{\bf 3}_H) + 6(\bar{\bf 3}_W,{\bf 3}_H)$ 
                   & 0 & 12 & 32    \\   \hline\hline
\end{tabular}
\]

\bigskip

\begin{center}
TABLE 1
\end{center}

\newpage

For the CW, CH, WH multiplets, the vector-like part in the second row,
like all the CC, WW, HH multiplets naturally acquire a mass $\sim M_U$.
Of the remaining chiral pieces, 45 of the 81 states are light being chiral
under the standard model gauge group and the remaining 36 also acquire
mass $\sim M_U$ under the symmetry breaking $3^{12} \rightarrow 3_C3_W3_H
\rightarrow 3_C2_L1_Y$ since they are vector-like under $3_C2_L1_Y$. 
Threshold effects occur when some of the heavy states lie below $M_U$.
We shall illustrate below, by examples, the magnitude of such effects.

There are 36 bifundamental scalars under $SU(3)^{12}$. These transform under the
$SU(3)_C \times SU(3)_W \times SU(3)_H$ subgroup 
and contribute to the $\Delta \beta_{3C,2L,Y}$ as shown
in Table 2. 

\[
\begin{tabular}{|l|c|c|c|}  \hline \hline
SCALAR MULTIPLET  & $\Delta \beta_C$ 
                  & $\Delta \beta_{2L}$ & $\Delta \beta_Y$   \\
\hline \hline
CC: $({\bf 3},\bar{\bf 3})_C$ 
                  & 1 & 0 & 0        \\  \hline
CW: $4({\bf 3}_C, \bar{\bf 3}_W) + (\bar{\bf 3}_C,{\bf 3}_W)$ 
                  & 5/2 & 5/2 & 10/3 \\ \hline
CH: $({\bf 3}_C, \bar{\bf 3}_H) + 4(\bar{\bf 3}_C,{\bf 3}_H)$ 
                  & 5/2 & 0 & 10/3   \\  \hline
WW: $4({\bf 3}, \bar{\bf 3})_W$ 
                  & 0 & 2 & 8/3      \\  \hline
HH: $4({\bf 3}, \bar{\bf 3})_H$   
                  & 0 & 0 & 8/3      \\  \hline
WH: $10({\bf 3}_W, \bar{\bf 3}_H) + 7(\bar{\bf 3}_W, {\bf 3}_H)$  
                  & 0 & 16 & 128/3   \\  \hline \hline
\end{tabular}
\]

\bigskip

\begin{center}
TABLE 2
\end{center}

\bigskip
\bigskip

All of the scalar representations are real under $3_C2_L1_Y$, indeed
under $SU(3)^{12}$, so all will naturally acquire a mass $\sim M_U$. One
$SU(2)_L$ doublet from the WH row of Table 2 must, however,
remain light as the standard Higgs doublet; this is the hierarchy problem.

\bigskip

Threshold effects are generally larger for fermions than for scalars,
as seen from Table 1 and 2. Let us therefore illustrate how fermion
masses below $M_U$ can effect the unification of $\alpha_{3C}$,
$\alpha_{2L}$ and $\alpha_Y$.

Without any threshold corrections, the consistent unification of teh
three couplings, $\alpha^{-1}_{3C, 2L, Y}$ is illustrated in Fig.\ 4.

Whether this unification survives threshold effects depends on the spectrum.
We illustrate this by Fig.\ 5-7. Fig.5 shows all the vector like
CH fermions at 2 TeV; Fig.\ 6 shows all the vector-like WH fermions at
2 TeV. In both cases, unification fails.
Fig.\ 7 shows all the vector-like CW fermions at 2 TeV; here, 
the unification is consistent at a higher scale $M_u \sim 5$ TeV. 
In all cases, $\alpha_{3C}(M_Z)$ and $\sin^2 \theta (M_Z)$ are 
at their experimental values.

Thus threshold effects are very significant because of the large
number of extra states and may spoil unification.  When $\Delta
\beta_{3C}, \Delta \beta_{2L}$ and $\Delta \beta_{Y}$ are comparable,
unification can remain consistent.  Similar results are obtained for
threshold effects from the scalar multiplets in Table 2.

\newpage

\bigskip
\bigskip

\noindent {\it Discussion}

\bigskip

The plots we have presented clarify the accuracy
of the predictions of this TeV unification scheme for
the precision values accurately measured at the Z-pole.
The predictivity is as accurate for $\sin^2 \theta$ as
it is for supersymmetric GUT models\cite{DG,DRW,ADF,ADFFL}. 
There is, in addition, an accurate prediction for $\alpha_3$
which is used merely as input in SusyGUT models.

At the same time, the accuracy of the predictions remains robust
if we allow the unification scale to vary
from about 2.5 TeV to 5 TeV. 

Threshold effects are large in some cases and may spoil 
unification, which depend on the spectrum of new states.

In conclusion, since this model ameliorates the GUT hierarchy
problem and naturally accommodates three families, it
provides a viable alternative to the widely-studied
GUT models which unify by logarithmic evolution
of couplings up to much higher scales.

\bigskip
\bigskip
\bigskip
\bigskip

\noindent {\it Acknowledgments}

\bigskip

\noindent This work was supported in part by the
US Department of Energy under
Grant No. DE-FG02-97ER-41036.

\newpage

\bigskip
\bigskip

\noindent \underline{\bf Figure Captions}

\bigskip

\noindent Fig. 1.

\noindent Plot of $\sin^2 \theta (M_U)$ versus $\alpha_3(M_U)/\alpha_2 (M_U)$
for various choices of $M_U$.

\bigskip

\noindent Fig.2.

\noindent Plot of $\sin^2 \theta(M_Z)$ versus $M_U$ in TeV, assuming
$\sin^2 \theta(M_U) = 1/4$ and $\alpha_3 / \alpha_2 (M_U) = 5/2$.

\bigskip

\noindent Fig.3.

\noindent Plot of $\alpha_3 (M_Z)$ versus $M_U$ in TeV, assuming
$\sin^2 \theta(M_U) = 1/4$ and $\alpha_3 / \alpha_2 (M_U) = 5/2$.

\bigskip

\noindent Fig.4.

\noindent Plot of $\alpha^{-1}_{3C}, (2/5)\alpha^{-1}_{2L},
(2/15)\alpha^{-1}_{Y}$ versus $E$(TeV) with no
threshold effects.

\bigskip

\noindent Fig.5.

\noindent Plot of $\alpha^{-1}_{3C}, (2/5)\alpha^{-1}_{2L},
(2/15)\alpha^{-1}_{Y}$ versus $E$(TeV) with 
all the vector-like CH fermions at 2 TeV.

\bigskip

\noindent Fig.6.

\noindent Plot of $\alpha^{-1}_{3C}, (2/5)\alpha^{-1}_{2L},
(2/15)\alpha^{-1}_{Y}$ versus $E$(TeV) with 
all the vector-like WH fermions at 2 TeV.

\bigskip

\noindent Fig.7.

\noindent Plot of $\alpha^{-1}_{3C}, (2/5)\alpha^{-1}_{2L},
(2/15)\alpha^{-1}_{Y}$ versus $E$(TeV) with 
all the vector-like CW fermions at 2 TeV.

\newpage

\begin{figure}

\begin{center}

\epsfxsize=7.0in
\ \epsfbox{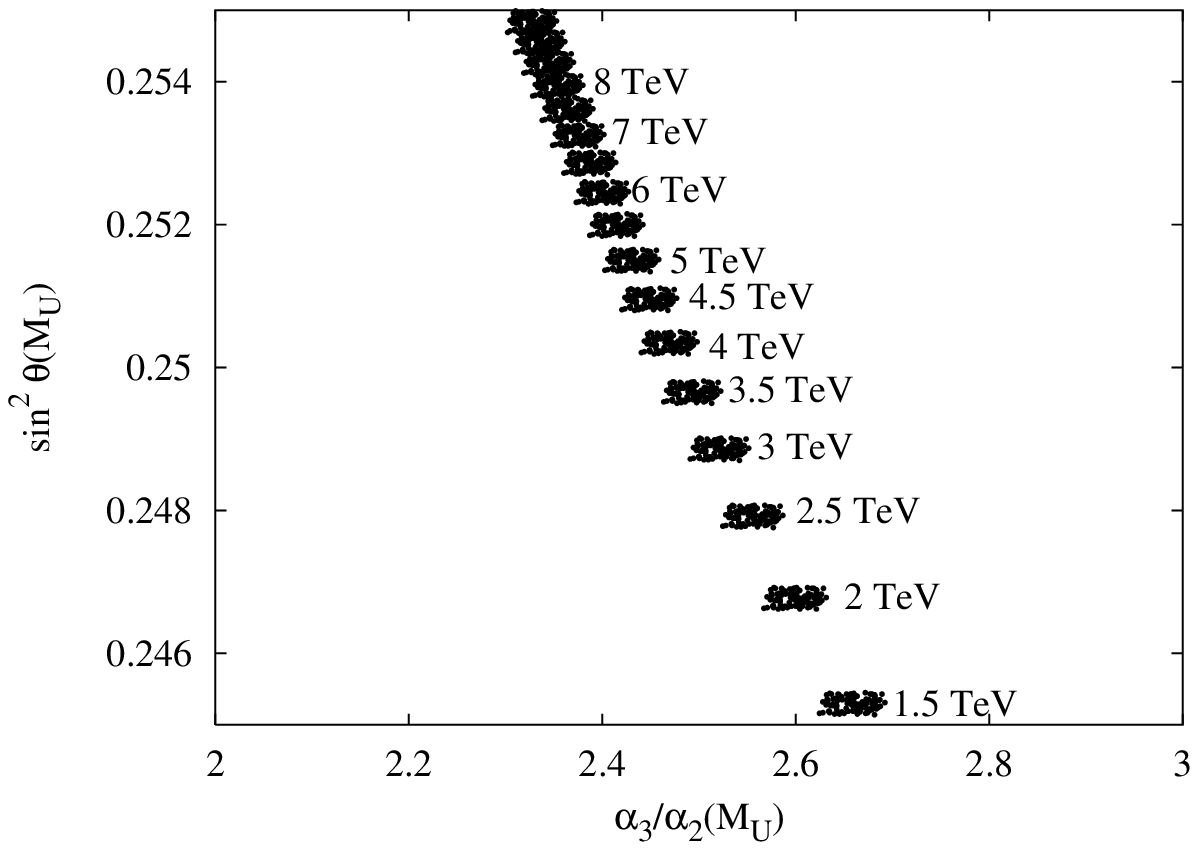}

\end{center}

\caption{}

\end{figure}

\newpage

\begin{figure}

\begin{center}

\epsfxsize=5.0in
\ \epsfbox{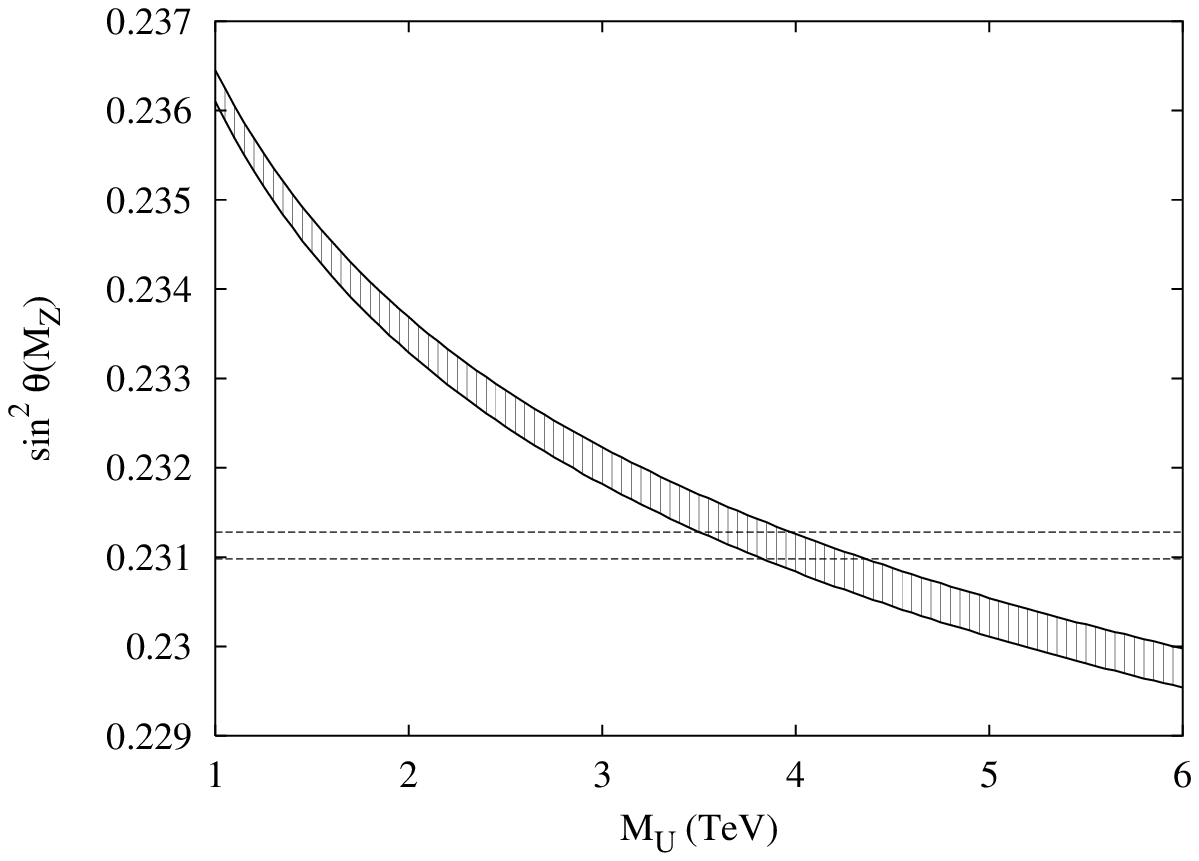}

\end{center}

\caption{}

 \end{figure}

\begin{figure}
\begin{center}
\epsfxsize=5.0in
\ \epsfbox{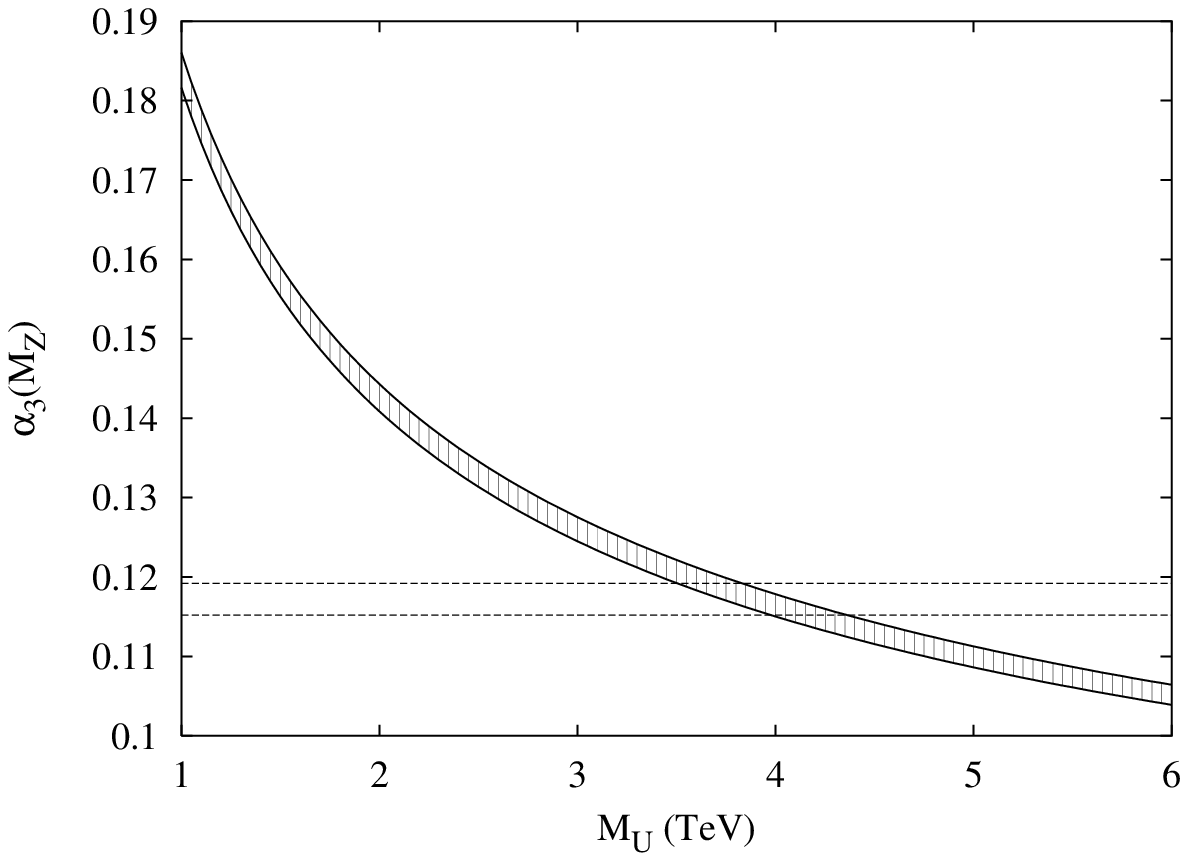}
\end{center}
\caption{}
\end{figure}

\begin{figure}
\begin{center}
\epsfxsize=5.0in
\ \epsfbox{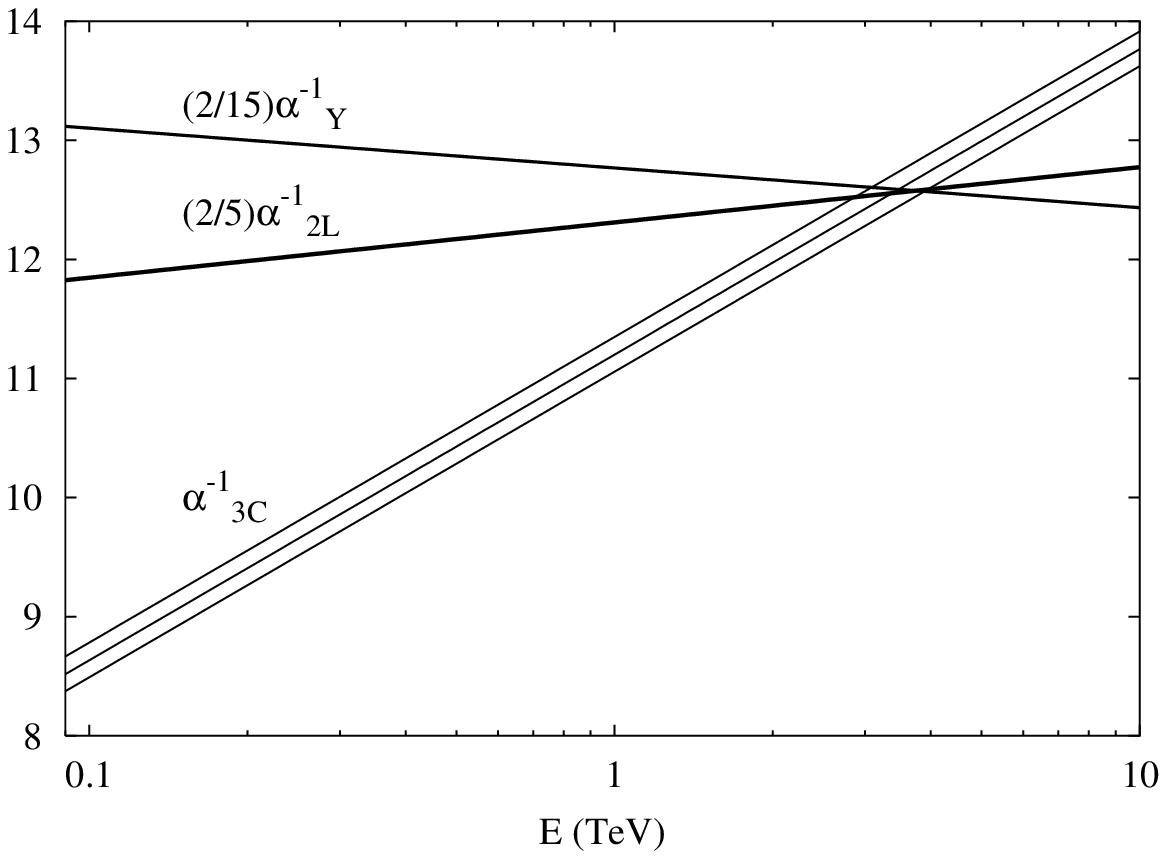}
\end{center}
\caption{}
\end{figure}

\begin{figure}
\begin{center}
\epsfxsize=5.0in
\ \epsfbox{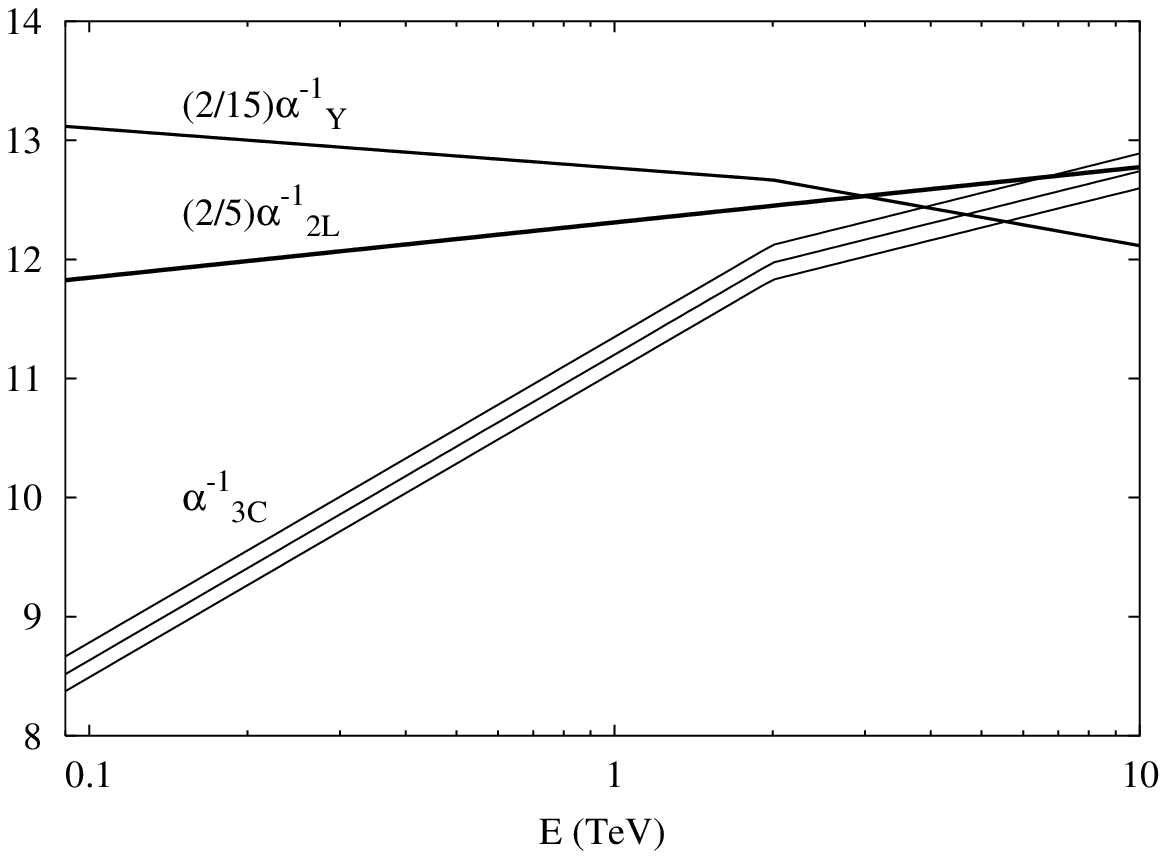}
\end{center}
\caption{}
\end{figure}

\begin{figure}
\begin{center}
\epsfxsize=5.0in
\ \epsfbox{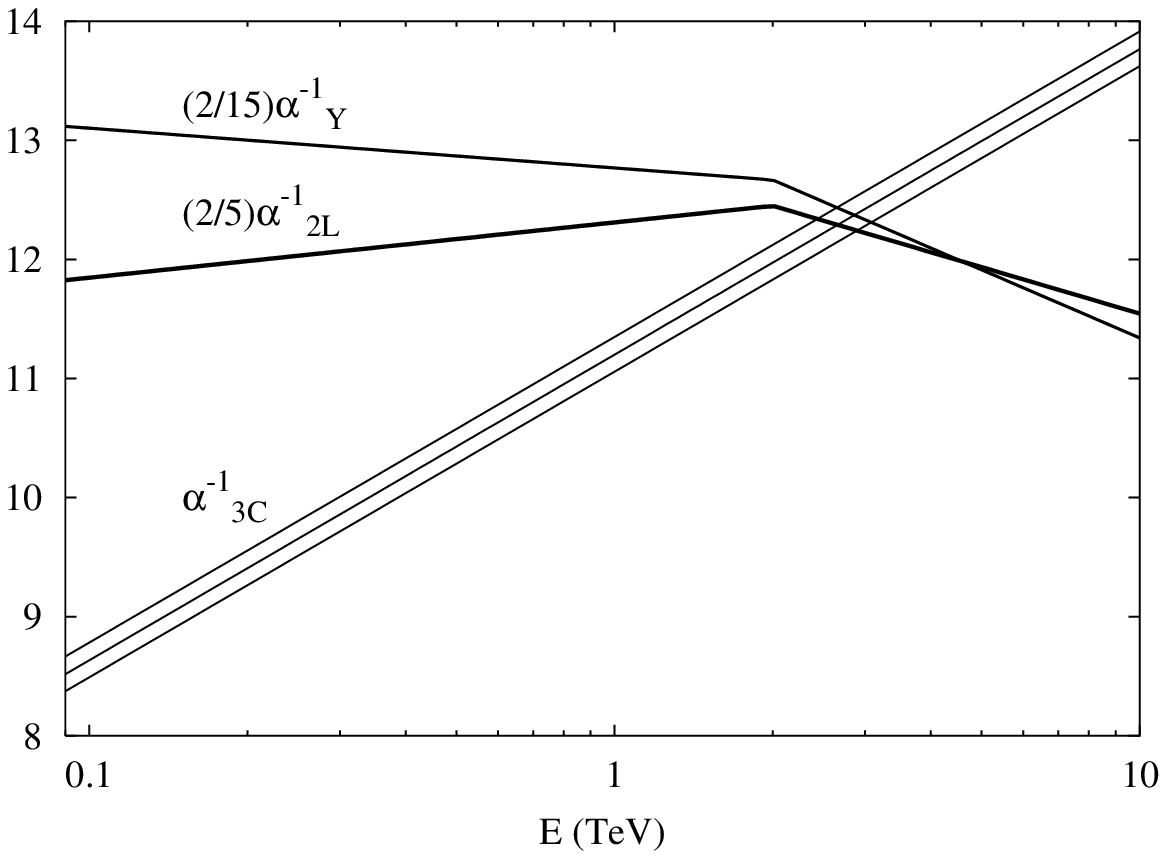}
\end{center}
\caption{}
\end{figure}

\begin{figure}
\begin{center}
\epsfxsize=5.0in
\ \epsfbox{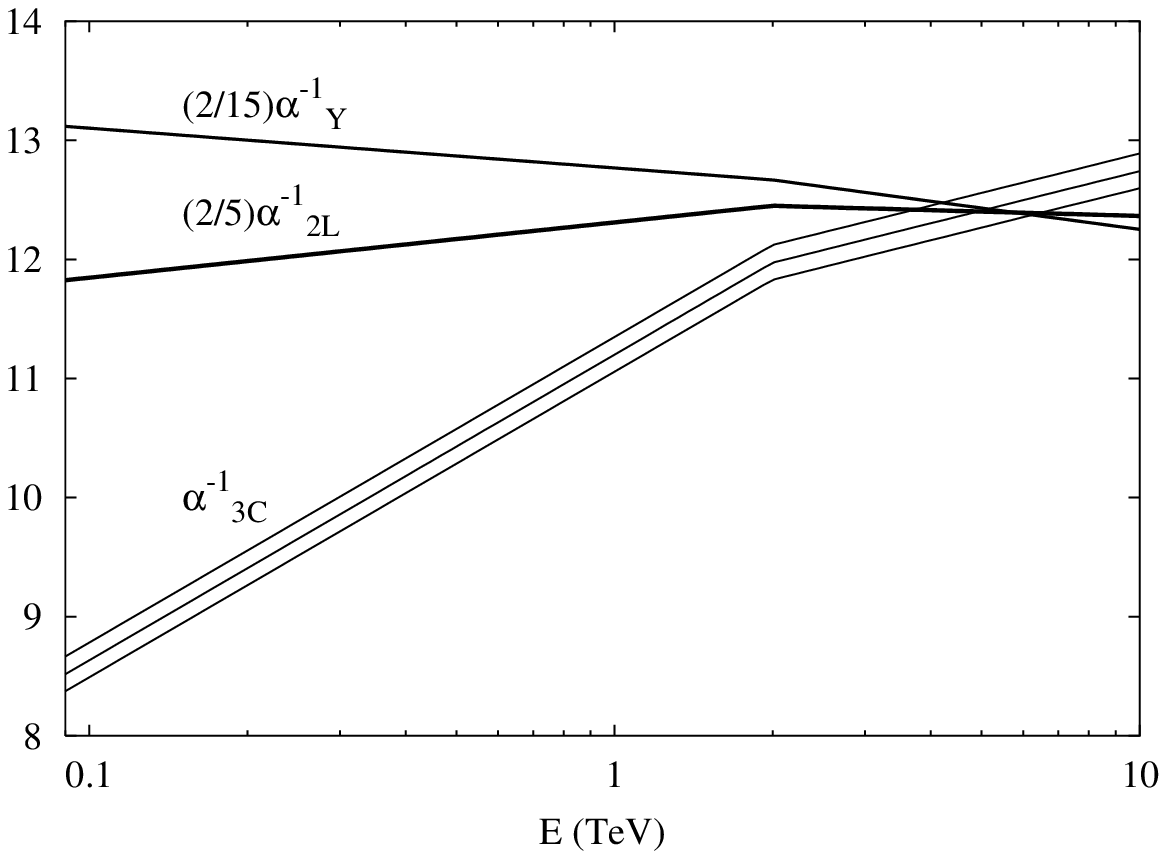}
\end{center}
\caption{}
\end{figure}

\end{document}